\documentclass[journal]{IEEEtran}

\usepackage{color}
\usepackage{graphicx, subfigure}
\usepackage{amsmath}
\usepackage{cite}
\usepackage{amsfonts}
\usepackage{multirow}
\usepackage{url}
\usepackage{threeparttable}
\usepackage{rotating}
\usepackage{hyperref}

\begin{document}
%TC:ignore
\title{AI for CSI Feedback Enhancement in 5G-Advanced}
%: An Industrial Perspective}
\author{\normalsize {Jiajia~Guo, \IEEEmembership{\normalsize {Graduate Student Member,~IEEE}},
Chao-Kai~Wen, \IEEEmembership{\normalsize {Senior Member,~IEEE}},\\
Shi~Jin, \IEEEmembership{\normalsize {Senior Member,~IEEE}},
and Xiao Li, \IEEEmembership{\normalsize {Member,~IEEE}}
}
\thanks{Jiajia Guo, Shi Jin, and Xiao Li are with the National Mobile Communications Research Laboratory, Southeast University, Nanjing, 210096, P. R. China (email: jiajiaguo@seu.edu.cn, jinshi@seu.edu.cn, li\_xiao@seu.edu.cn).}
\thanks{Chao-Kai~Wen is with the Institute of Communications Engineering, National Sun Yat-sen University, Kaohsiung 80424, Taiwan (e-mail: chaokai.wen@mail.nsysu.edu.tw).}
}

\maketitle

\begin{abstract}
The 3rd Generation Partnership Project started the study of Release 18 in 2021.
Artificial intelligence (AI)-native air interface is one of the key features of Release 18, where AI for channel state information (CSI) feedback enhancement is selected as the representative use case.
{{This article provides an overview of AI for CSI feedback enhancement in 5G-Advanced.
Several representative non-AI and AI-enabled CSI feedback frameworks are first introduced and compared.
Then, the standardization of AI for CSI feedback enhancement in 5G-advanced is presented in detail.
First, the scope of the AI for CSI feedback enhancement in 5G-Advanced is presented and discussed.
Then, the main challenges and open problems in the standardization of AI for CSI feedback enhancement, especially focusing on performance evaluation and the design of new protocols for AI-enabled CSI feedback, are identified and discussed.
This article provides a guideline for the standardization study of AI-based CSI feedback enhancement.
}}

\end{abstract}
%TC:endignore

\IEEEpeerreviewmaketitle

\section{Introduction}

\IEEEPARstart{I}{n} June 2022, the 3rd Generation Partnership Project (3GPP) completed the latest release of the fifth generation (5G) cellular networks, namely, Release 17, which is the last version of the first stage of 5G evolution.
With the functional freeze of Release 17, the 3GPP started planning Release 18, the first release of the second stage of 5G evolution, namely, 5G-Advanced \cite{9795045,lin2022overview}.
%Forty-one work packages have been proved for Release 18 in 2021 and 2022.
Artificial intelligence (AI) is regarded as one of the key features of 5G-Advanced and sixth generation (6G) cellular networks, and it will trigger a paradigm shift and lay a strong foundation for 5G-Advanced and 6G \cite{9795045,lin2022overview,9446676}.

{{Unlike the conventional methods supported by domain knowledge and rigorous theoretical proofs, AI-based methods, including deep learning and reinforcement learning, can automatically learn or extract features from a training dataset via neural networks (NNs).
To efficiently learn the features from the dataset, considerable NN architectures, such as dense, convolutional, and recurrent NNs, are developed.
Inspired by AI's great success in computer vision, it has been introduced to wireless communications in the past few years \cite{9446676}.}}
AI-enabled radio access networks were investigated in Release 17 \cite{lin2021fueling}.
However, the AI-native air interface is excluded in the study scope of Release 17.
In December 2021, a new study item on AI-native air interface was proposed in Release 18 \cite{3gpp599}.
This study item aims to explore the benefits of integrating AI into the air interface, including performance improvement and complexity reduction.
In particular, this study item focuses on assessing the performance of the AI-native air interface compared with conventional methods and its specification impacts, establishing a common AI framework for the AI-native air interface and identifying the requirement in an AI-native air interface.
The use-case-driven approach is employed for this item, and three representative use cases, including channel state information (CSI) feedback enhancement, beam management, and position accuracy enhancement, are selected to investigate the performance, complexity, and potential specification impacts of the AI-native air interface \cite{3gpp599}.

Herein, we focus on the first use case, that is, AI for CSI feedback enhancement.
Massive multiple-input and multiple-output (MIMO), which equips the base station (BS) with many antennas, can considerably improve the system performance.
However, the benefit of massive MIMO is based on the knowledge of downlink CSI.
In frequency division duplexing (FDD) massive MIMO systems, the user equipment (UE) must feedback the downlink CSI to the BS through the uplink because of the lack of channel reciprocity.
The feedback overhead is substantial due to the high dimension of the CSI in massive MIMO systems. 
Recently, AI has been introduced to CSI feedback by \cite{wen2018deep}.
AI-enabled CSI feedback learns to automatedly compress and reconstruct CSI and considerably improves the feedback accuracy compared with codebook- and compressive sensing (CS)-based feedback algorithms{{\cite{guo2022overview}}}.
AI-enabled CSI feedback has been selected as a use case in \cite{3gpp599} because of its excellent performance, as supported by approximately 50 companies.
However, many problems should be clarified and solved before deploying AI-enabled CSI feedback to practical systems.
{{
For example, most of the existing works are based on simulation and cannot be fairly compared with conventional feedback methods \cite{3gppType} in accuracy, complexity, and generalization.
%Their performances in practical systems are unclear.
Moreover, the standardization of AI-native air interface technology has not been considered and discussed by the 3GPP, and developing a protocol for autoencoder-based CSI feedback is a challenge.

For an overview of AI-enabled CSI feedback, we refer the interested readers to \cite{guo2022overview}.
In contrast to \cite{guo2022overview}, this paper introduces AI-enabled CSI feedback enhancement from the industrial perspective and the challenges and open problems in the deployment of AI-enabled CSI feedback, especially in standardization of 5G-Advanced.
The rest of the article is organized as follows.
Section \ref{AIvsNonAI} introduces and compares the representative non-AI and AI-enabled CSI feedback frameworks.
Then, the scope of AI for CSI feedback enhancement in 5G-Advanced is briefly presented and discussed in Section \ref{considerations}.
The main considerations in the standardization of AI for CSI feedback enhancement in 5G-Advanced are then provided and discussed in Section \ref{considerations}.
On the one hand, the performance gains, including feedback accuracy, feedback overhead, and algorithm complexity, should be fairly evaluated and compared with the existing feedback standard in 5G new radio (NR).  
On the other hand, several new challenges to standardization by AI-enabled CSI feedback enhancement should be clarified and addressed.
Finally, Section \ref{future} concludes this article.

\section{Representative Non-AI and AI-enabled CSI Feedback Frameworks}
\label{AIvsNonAI}
%Codebook- and CS-based CSI feedback strategies are introduced to reduce feedback overhead.
%However, the feedback accuracy and algorithm complexity of these strategies cannot satisfy the system requirement.
%, thereby impeding the development of the FDD mode in 5G and beyond.
In this section, non-AI feedback frameworks (including codebook- and CS-based feedback) and their main shortcomings are introduced.
Then, three representative AI-enabled CSI feedback frameworks are presented, respectively.
Next, the non-AI and AI-enabled frameworks are compared in terms of feedback accuracy and overhead.
Finally, the advantages of AI-enabled over non-AI frameworks and why AI is the CSI feedback direction are given.

\subsection{Non-AI CSI Feedback Frameworks}
The existing non-AI feedback strategies can be divided into two main kinds: codebook- and CS-based frameworks.

\subsubsection{Codebook-based CSI Feedback}
\label{cedebook}
Codebook-based feedback is adopted in existing communication systems \cite{3gppType}.
In this framework, the UE searches the nearest codeword in a predefined codebook, which is shared by the UE and the BS.
Then, the UE feedbacks the index of the selected codeword to the BS.
The BS obtains the corresponding codeword by looking up the codebook.

This feedback faces two main challenges in massive MIMO systems.
On the one hand, the codebook resolution and feedback accuracy are improved with the codebook size, that is, the feedback bit number.
In massive MIMO systems, the channel dimension is high, and many bits are needed to improve the feedback accuracy.
On the other hand, the algorithm complexity increases with the codebook size.
%The simplified codeword search algorithms unavoidably reduce the feedback accuracy. 
%Therefore, the accuracy and algorithm complexity of the codebook-based feedback framework cannot fully meet the requirements of 5G-Advanced and beyond.

\subsubsection{CS-based CSI Feedback}
In CS theory, the original signal can be exactly reconstructed from the compressed signal if sparsity exists.
CS-based CSI feedback has been widely studied in the past 10 years.
The basic assumption of this type of feedback is the CSI sparsity in a certain domain, such as the angular and delay domain.
On the basis of this assumption, the downlink CSI is compressed by a sensing matrix and then reconstructed by some CS algorithms.

The main concerns of the CS-based feedback are the sparsity assumption and the complexity of the reconstruction algorithms.
First, the sparsity characteristic cannot fully describe the structure of the CSI matrix.
Second, the reconstruction problem in CS is usually solved by some iterative algorithms.
The time complexity of these reconstruction algorithms is extremely high to meet the requirement of practical systems.

}}

\begin{figure}[t]
	\centering 
	\includegraphics[width=0.4\textwidth]{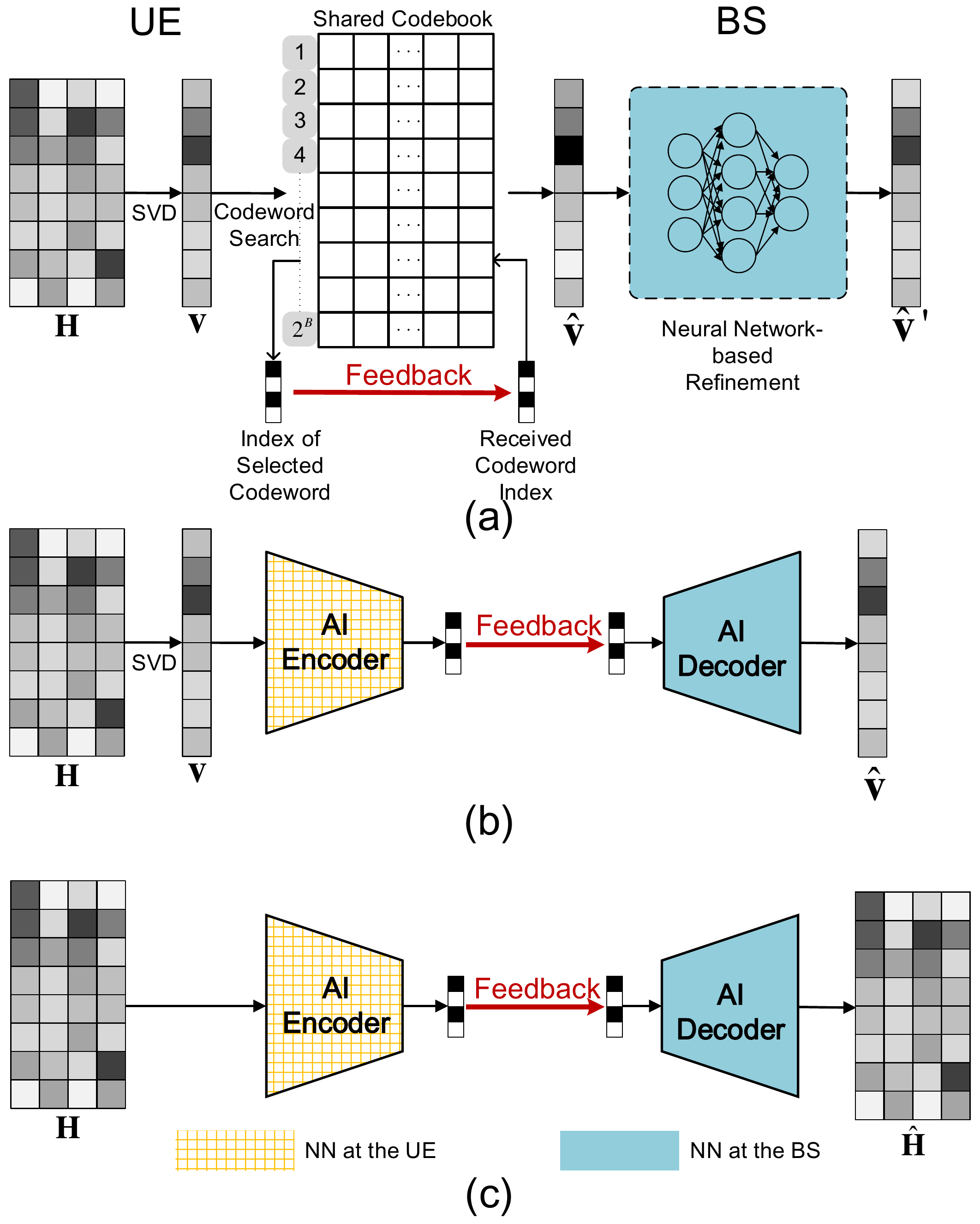}
	\caption{\label{ThreeFramework}{Three representative frameworks of AI for CSI feedback}, including (a) one-sided refinement for implicit CSI feedback, (b) two-sided enhancement for implicit CSI feedback, and (c) two-sided enhancement for explicit CSI feedback.}  
  \vspace{-0.75cm}
\end{figure}

\subsection{AI-enabled CSI Feedback Frameworks}

Many novel AI-enabled technologies have been introduced and studied in the past few years.
However, these technologies cannot be immediately deployed in practical systems.
The technologies that need no/few modifications for application in existing systems are prioritized in 5G-Advanced.
Meanwhile, the technologies that need to totally change the existing systems, such as autoencoder-based end-to-end communications, will be studied in 6G and beyond.
Reference \cite{9446676} divided the technologies in AI-native air interface into three types: replacing single communication modules, designing several modules jointly with AI, and totally changing the existing block-by-block communication framework.
On the basis of \cite{9446676}, the existing frameworks of AI for CSI feedback enhancement are divided into three types, namely, one-sided refinement and two-sided enhancement for implicit CSI feedback and two-sided enhancement for explicit CSI feedback, as shown in Fig. \ref{ThreeFramework}.

\subsubsection{AI-enabled One-sided Refinement for Implicit CSI Feedback}
\label{f1}
Fig. \ref{ThreeFramework}(a) shows the framework of the AI-enabled one-sided refinement for CSI feedback, which need not change the existing feedback framework and is easy to deploy \cite{9789120}.
Once the BS is installed, the environment around the BS remains stable for a long time.
Given that wireless channels greatly depend on the propagation environment, the channel of a certain cell shows certain characteristics that can be regarded as environmental knowledge.
AI is a good tool that extracts and utilizes environmental knowledge to help CSI feedback via end-to-end learning.

Specifically, the UE first performs singular value decomposition on the downlink CSI ($\bf H$) to produce the eigenvectors, that is, the precoding matrix ($\bf V$) \cite{9662381}.
The UE feedbacks the precoding matrix with the codebook shared by the BS.
Upon receiving the index of the selected precoding codeword, namely, the precoding matrix indicator (PMI), the BS selects the corresponding codeword through the shared codebook.
Then, the pretrained NN refines the obtained codeword.
The accuracy of the refined precoding matrix is considerably improved with the help of environmental knowledge.
The training and inference operations of NNs are conducted at the BS without the participation of the UE.
In the training phase, the input and output of the NN are the selected channel codeword ($\hat {\bf v}$) and the perfect precoding matrix ($\bf v$), respectively.

{
In addition, the NN module can be also added at the UE side.
As mentioned in Section \ref{cedebook}, the complexity of the codeword search is the major challenge in codebook-based CSI feedback.
Therefore, adding an NN module at the UE side to help codeword search is a promising approach that is worth exploring.
}

\subsubsection{Autoencoder-based Two-sided Enhancement for Implicit CSI Feedback}
Fig. \ref{ThreeFramework}(b) shows the autoencoder-based two-sided framework for implicit CSI feedback enhancement \cite{9662381}.
Unlike that in Fig. \ref{ThreeFramework}(a), this framework must change the existing feedback strategy.
Inspired by autoencoder-based image compression, an NN-based encoder is adopted at the UE to compress and quantize the generated precoding matrix.
We can regard the generated bitstream in this framework as the PMI in the existing codebook-based feedback strategy.
After obtaining the feedback bitstream, the NN-based decoder reconstructs the original precoding matrix.

\subsubsection{Autoencoder-based Two-sided Enhancement for Explicit CSI Feedback}
\label{3rdFrame}
Fig. \ref{ThreeFramework}(c) shows the autoencoder-based two-sided framework for explicit CSI feedback enhancement \cite{wen2018deep}.
In contrast with the above-mentioned frameworks, this third framework feedbacks the entire downlink CSI, $\bf H$, to the BS via an NN-based encoder.
Then, the decoder at the BS reconstructs the original CSI on the basis of the received bitstream.
Given that all information of the downlink CSI is fed back, the feedback overhead of this framework is larger than that of the second framework.
This framework is the most widely studied in academia.

\subsubsection{Comparison among Three Frameworks}
The three frameworks presented above have been introduced to CSI feedback enhancement, and their main differences include their feedback information form and effects on the air interface.
Implicit feedback is adopted in the first two frameworks, while the last one adopts an explicit feedback strategy.
Implicit feedback is used in 5G NR.

Explicit feedback may bring more performance gains than implicit feedback.
However, the utilization of the downlink CSI in the entire system should be changed.
The CSI utilization in practical systems is based on the partial CSI.
In addition, AI-enabled explicit feedback cannot be directly compared with the feedback strategy adopted in practical systems.
AI-based implicit feedback enhancement frameworks can be evaluated and compared with the Type II codebook. 

The first framework, which refines the obtained implicit CSI at the BS, adopts the existing feedback strategy.
The last two frameworks need to replace the original codebook-based coding and decoding with the NN-based encoder and decoder, respectively.
Specifically, the existing feedback standard should be completely changed.
Accordingly, we believe that these three frameworks will be deployed in practical systems in different stages.
The first framework is plug-and-play and can be embedded into the existing BS without standardization.
The second framework, which feedbacks the implicit CSI with an autoencoder, will be introduced in 5G-Advanced.
The last framework, which completely changes the CSI feedback and utilization strategy, will be deployed in 6G and beyond.

\begin{table}[t]
\caption{\label{Table1}{{GCS comparison between the second AI-enabled framework and Type II codebook \cite{9662381}}}}
\centering
\setlength{\tabcolsep}{5mm}{
\begin{tabular}{c|c|c}
\hline \hline 
Methods                      & Feedback Bit Number & GCS   \\
\hline \hline 
Type II Codebook             & 300                  & 0.9042 \\ \hline
\multirow{6}{*}{AI-enabled feedback} & 78                   & 0.8507 \\
                             & 104                   & 0.8721 \\
                             & \textbf{156}                   & \textbf{0.9044} \\
                             & 208                  & 0.9221 \\
                             & 312                  & 0.9448 \\
                             & 416                  & 0.9574 \\
                             \hline \hline 
\end{tabular}
}
 \vspace{-0.75cm}
\end{table}

{
\subsection{Performance Comparison between Non-AI and AI-enabled CSI Feedback Frameworks}
%Some simulation results and comparisons between the no-AI and AI-enabled CSI feedback methods are provided in this subsection.
%Then, the advantages of AI-enabled CSI feedback over non-AI feedback are discussed.

Implicit feedback is adopted in 5G NR.
Therefore, we first compare the AI-enabled implicit CSI feedback with the Type II codebook adopted in 5G NR.
Table \ref{Table1} shows the generalized cosine similarity (GCS) performance comparison between the second AI-enabled framework and the Type II codebook.
The simulation settings satisfy the simulation requirement of the 3GPP, and the simulation details can be found in \cite{9662381}.
The table indicates that the AI-enabled implicit feedback still outperforms the Type II codebook when the feedback overhead is approximately reduced by 50\%.

As mentioned in Section \ref{3rdFrame}, the third framework, that is, the autoencoder-based explicit CSI feedback, is the most widely studied in academia.
Therefore, this type of feedback has been compared with conventional CS-based feedback algorithms, as shown in \cite{wen2018deep}.
Table \ref{NMSEcom} shows the normalized mean-squared error (NMSE) performance of the CS-based algorithms and the AI-enabled methods, including CsiNet \cite{wen2018deep} and TransNet \cite{9705497}.
TransNet adopts more advanced NNs than the vanilla CsiNet.
Table \ref{NMSEcom} shows that the conventional algorithms perform much worse than the AI-enabled methods under all compression ratios.
The AI-enabled methods still perform well when the conventional algorithms do not work at all.
For example, if the compression ratio is set to 1/32 and the indoor scenario is considered, the best NMSE achieved by the conventional algorithms is only $-$0.27 dB, while that of TransNet reaches $-$10.49 dB.

Except for the advantage in the feedback accuracy, the inference speed of AI-enabled CSI feedback is higher than that of the conventional algorithms.
For example, CsiNet performed over 50 times faster than the CS-based algorithms in \cite{wen2018deep}.
The inference time of the AI-enabled implicit CSI feedback in \cite{9662381} is below 1 ms.

%The performance advantages of AI-enabled CSI feedback in accuracy and complexity are easily demonstrated through the above comparisons, indicating the great potential of AI-enabled CSI feedback in 5G-Advanced and beyond.

\begin{table}
	\centering
	\caption{\label{NMSEcom}{NMSE {\rm dB} performance comparison of different explicit CSI feedback algorithms \cite{wen2018deep,9705497} }}
	\label{table:result}
		\begin{tabular}{cc|cc}
			\hline  \hline
			compression ratio & Methods & Indoor      & Outdoor    \\  \hline\hline
			& LASSO                                         & $-$7.59                    & $-$5.08                   \\
			& BM3D-AMP                                      & $-$4.33                  & $-$1.33                 \\
			1/4                       & TVAL3                                         & $-$14.87                  & $-$6.90               \\
			& CsiNet                                        & {$-$17.36} &   {$-$8.75}  \\ 
			& TransNet                                        & \textbf{$-$32.38} &  \textbf{$-$14.86}  \\
			
			\hline
			& LASSO                                         & $-$2.72              & $-$1.01           \\
			& BM3D-AMP                                      & 0.26                  & 0.55                \\
			1/16                      & TVAL3                                         & $-$2.61                & $-$0.43               \\
			& CsiNet                                        & {$-$8.65}   & {$-$4.51}  \\
			& TransNet                                        & \textbf{$-$15.00} &  \textbf{$-$7.82}  \\
			\hline
			& LASSO                                         & $-$1.03             & $-$0.24           \\
			& BM3D-AMP                                      & 24.72          & 22.66           \\
			1/32                      & TVAL3                                         & $-$0.27             & 0.46                \\
			& CsiNet                                        & {$-$6.24}  &  {$-$2.81}\\ 
			& TransNet                                        & \textbf{$-$10.49} &  \textbf{$-$4.13}  \\
			\hline
			& LASSO                                         & $-$0.14                 & $-$0.06              \\
			& BM3D-AMP                                      & 0.22               & 25.45                  \\
			1/64                      & TVAL3                                         & 0.63                   & 0.76                 \\
			& CsiNet                                        & {$-$5.84}  &  {$-$1.93} \\
			& TransNet                                        & \textbf{$-$6.08} &  \textbf{$-$2.62}  \\
			\hline  \hline
		\end{tabular}
\end{table}

\subsection{Reason why CSI Feedback would Go towards AI}
\label{reasonWhy}

Interpretability is the main challenge of data-driven algorithms, which lack rigorous theoretical proofs, and AI-enabled CSI feedback is no exception.
The key to AI is learning and utilizing data distribution.
For a certain cell or channel model, the CSI distribution remains stable.
The distribution can be viewed as a kind of environmental knowledge because CSI is highly dependent on the propagation environment.
The NNs for CSI feedback can learn this knowledge via end-to-end training with many CSI samples.
As shown in Fig. \ref{envi}, the encoder at the UE compresses the CSI based on the learned knowledge.
Accordingly, the decoder reconstructs the CSI based not only on the received compressed CSI but also on environmental knowledge.
The compression at the UE is highly efficient, and the reconstruction is more accurate due to extra knowledge.

\begin{figure}[t]
	\centering 
	\includegraphics[width=0.45\textwidth]{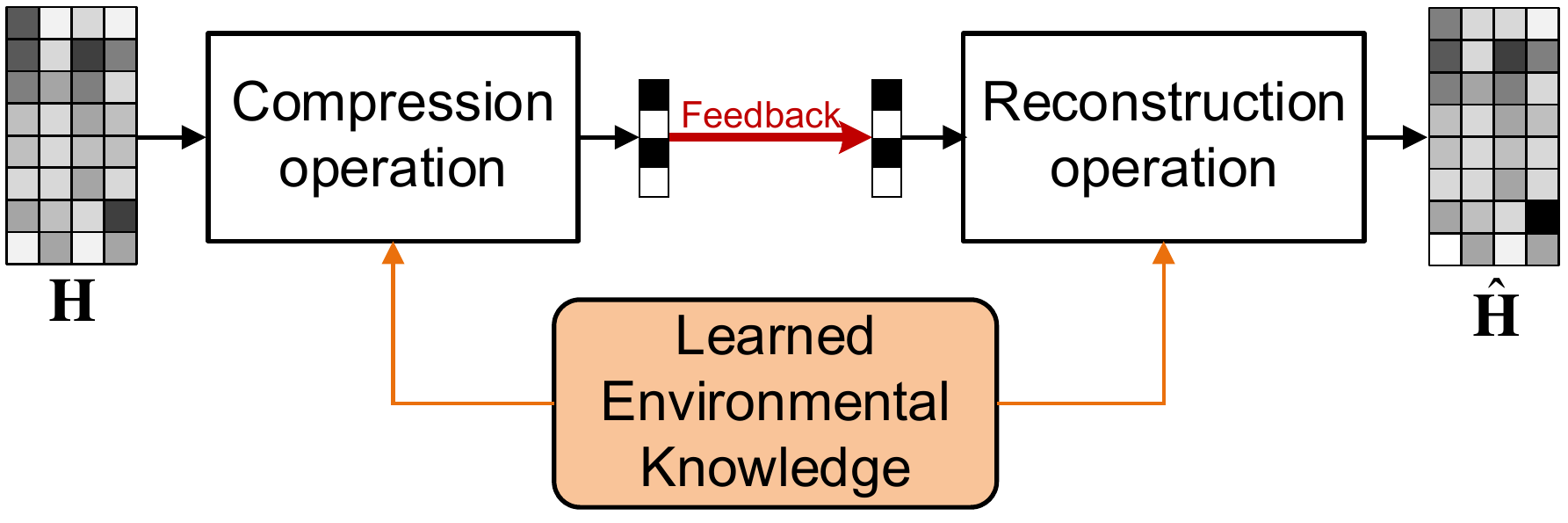}
	\caption{\label{envi}Illustration of CSI compression and reconstruction with the aid of learned environmental knowledge}  
\end{figure}

Conventional non-AI algorithms also try to introduce this knowledge to CSI compression and reconstruction.
The knowledge is on top of the hand-engineered distribution or channel structure.
The mismatch between the original CSI and assumed distributions is unavoidable, leading to a considerable drop in feedback accuracy.
However, the knowledge used in AI-enabled feedback is learned from the datasets and can exactly describe the propagation environment.
Therefore, AI-enabled CSI feedback will play an essential role in 5G-Advanced and beyond.

%Although AI-enabled algorithms have high complexity, the inference speed can be considerably accelerated by graphics processing units (GPUs) because of the parallel architecture.
%When AI-enabled communication algorithms are deployed, some specific hardware accelerators further reduce the inference time.

}

\section{Standardization of AI for CSI Feedback Enhancement in 5G-Advanced}
\label{considerations}
The transceivers of different companies can use different technologies and need not be shared.
However, the downlink CSI feedback in the air interface should be kept the same for all UE and BSs, and the key modules in the CSI feedback must be standardized.
Therefore, many problems should be considered and clarified.
{{
In this section, the scope of AI for CSI feedback enhancement is presented.
Then, the main challenges and open problems during the standardization are presented.
The 3GPP focuses on standardizing key frameworks, which all companies must follow.
The detailed techniques need not be standardized, and the companies can use their own techniques.
Therefore, on the basis of the scope and goal of this paper, we focus on standardization and do not discuss the techniques in detail.
}}

\subsection{Scope of AI for CSI Feedback Enhancement in 5G-Advanced}

The main challenges of the CSI feedback include the substantial uplink bandwidth, the low reconstruction accuracy at the BS, and channel aging.
Therefore, the main scope of AI for CSI feedback enhancement in the AI-native air interface study item of the 3GPP Release 18 \cite{3gpp599} involves overhead reduction, accuracy improvement, and channel prediction.

\subsubsection{Overhead Reduction}
\label{scope1}
The feedback accuracy of codebook-based CSI feedback improves but the feedback overheads increase.
In a 5G NR, the Type II codebook is adopted in the multi-UE scenario, where a highly accurate CSI is needed to reduce the interference among different UE compared with the Type I codebook adopted in the single-UE scenario \cite{3gppType}.
For example, the feedback bit number is approximately 300 when the respective numbers of antennas of the BS and the UE are 32 and 4, and the sub-band number is 13 (additional details can be found in \cite{9662381}).
This feedback overhead consumes substantial resources of the uplink control channel.
Therefore, AI-enabled CSI feedback is expected to reduce the feedback overhead without a decline in accuracy. 

\subsubsection{Accuracy Improvement}
\label{scope2}
The downlink performance greatly depends on the quality of the downlink CSI obtained at the BS.
However, the feedback accuracy in practical systems does not meet the requirement of future communication systems.
GCS is widely used to evaluate the CSI feedback performance.
According to \cite{9662381}, the GCS of the Type I codebook adopted in the single-UE scenario is only 0.76.
The accuracy of the codebook-based feedback is low, unavoidably leading to a throughput drop in downlink transmission.
Therefore, an AI-enabled feedback method that considerably improves the feedback accuracy without overhead increase must be developed.

\subsubsection{Channel Prediction}
\label{scope3}
The speed of the UE mobility is high in some cases.
For example, the speed of a high-speed train is up to 350 km/h, where the propagation environment rapidly changes.
In these high-speed cases, channel aging seriously negatively affects the downlink transmission performance even when the downlink CSI is perfectly fed back to the BS.
{Moreover, feedback is no longer needed if the difference between the estimated CSI at the UE and the predicted CSI at the BS is acceptable.}
Therefore, channel prediction should be considered when AI is introduced to CSI feedback enhancement.

{
\subsection{Main Challenges and Open Problems during Standardization}
\label{mainCha}
%AI-native air interface is a new topic for the 3GPP.
Many challenges and problems must be articulated and solved during the standardization phase. 
%First, AI-enabled CSI feedback is only worth deploying if it truly outperforms the existing feedback algorithms.
%Therefore, 
First, the performance of AI-enabled CSI feedback must be evaluated and fairly compared with that of the existing methods.
Sections \ref{mainCha}-(1-3) introduce the performance evaluation of AI for CSI feedback enhancement in 5G-Advanced.
%, including accuracy, complexity, and generalization.
Then, AI-enabled CSI feedback creates new requirements for standardization, and the corresponding protocols must be carefully designed for deployment.
Sections \ref{mainCha}-(4-7) introduce the collaboration between the UE and the BS, the shared information between different companies, the joint design with channel prediction, and the combination with reciprocity-based feedback.
}

\subsubsection{Accuracy Evaluation of the AI-enabled CSI Feedback}
\label{PerformanceEvaluation}
Most of the existing studies on AI for CSI feedback enhancement were conducted on the COST 2100 channel dataset shared by \cite{wen2018deep}.
This channel dataset is simple and far from the practical systems, and thus cannot be used to compare the AI-based feedback with the existing systems.
The accuracy evaluation of CSI feedback should be fair and easily show the potential of AI-based CSI feedback enhancement.

First, system-level simulation is preferred over link-level simulation.
To the best of our knowledge, link-level simulation was adopted in nearly all of the existing work.
Consequently, the performance of AI for CSI feedback enhancement in systems is unclear.
System-level simulation is beneficial in achieving a comprehensive performance in throughput in a single- or multi-cell perspective.
Moreover, system-level performance is widely adopted in 5G NR MIMO enhancement and near the complicated practical systems.
The performance of AI-enabled CSI feedback can be easily compared with the existing 5G NR technology.
Given that system-level simulation is time-consuming, link-level simulation can be optionally adopted to evaluate the novel works on CSI feedback.

Second, as previously mentioned, the COST 2100 channel dataset is used to evaluate the CSI feedback.
However, this channel model is not adopted in 5G NR and cannot be used to compare the AI-enabled methods with the Type I and Type II codebooks in 5G NR.
Consequently, the agreed 3GPP channel models should be used, such as TR 38.901.
Moreover, the channel settings adopted in the CSI feedback enhancements of Releases 16 and 17 can be directly used, facilitating the comparison of AI-enabled CSI feedback with the latest codebook in 5G NR, that is, the enhanced Type II codebook.
%The indoor and the outdoor scenarios, where the mobility speeds of the UEs are 3 and 30 km/h, need to be considered together.
In addition, the downlink channel at the UE was assumed perfect, that is, an ideal downlink channel estimation, in most works.
The realistic channel estimation must be considered during feedback to evaluate its effects on AI-based feedback.

Finally, NMSE and GCS are the most widely used key performance indicators (KPIs) in AI-enabled CSI feedback work \cite{wen2018deep,guo2022overview}.
Given that system-level simulation is preferred, the KPIs consist of two types: intermediate and eventual KPIs.
The first type is adopted to compare the feedback accuracy of different NN models.
NMSE, GCS, and square GCS can be used in this part.
The second type evaluates the performance of the entire system and is used to compare the AI methods with the conventional methods.
The throughput and the overhead can be used in this case.

Overall, the accuracy evaluation of AI-enabled CSI feedback should be fair to show the potential of AI methods compared with non-AI methods under the agreed 3GPP assumption \cite{xiaodong}.

\subsubsection{Complexity of the AI-enabled CSI Feedback}
\label{ComplexityOf}
Communication algorithms have a high requirement in inference latency and computational complexity \cite{9770094}.
Although AI-enabled CSI feedback promises to outperform the conventional algorithms, its computational complexity is a major challenge because of the following reasons.
First, the performance of AI-enabled methods is improved by stacking many NN layers or widening the NN layers.
Performance improvement is achieved with a considerable increase in NN complexity.
Second, all simulations are conducted on a computer/workstation equipped with several powerful graphics processing units (GPUs), which have more computational power than the computing units of the UE.
Finally, the NN weights must be stored, thereby occupying memory storage in the UE.
Therefore, the complexity should be evaluated and reduced during the deployment of AI-enabled CSI feedback.

  \begin{figure*}[t]
	\centering 
	\includegraphics[width=0.75\textwidth]{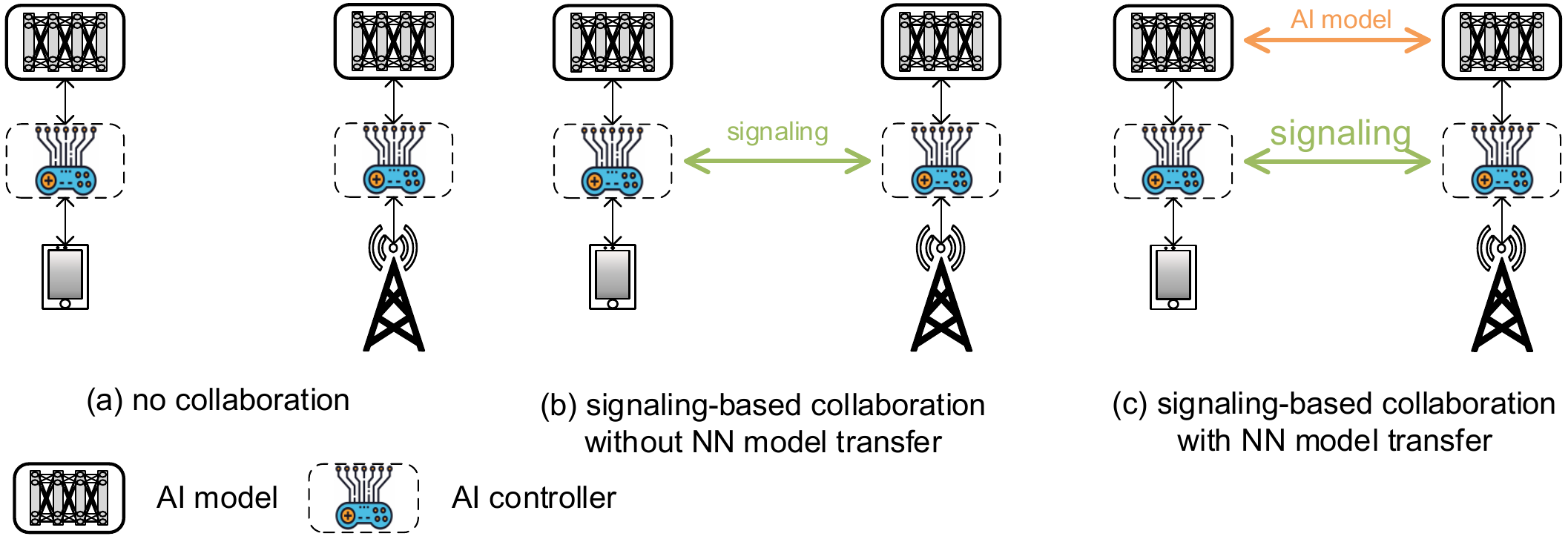}
	\caption{\label{Collaboration}Three collaborations between the UE and the BS of the AI-enabled CSI feedback.}  
  \vspace{-0.75cm}
\end{figure*}

A major problem is evaluating the complexity of AI-based CSI feedback.
The number of floating-point operations (FLOPs) is widely adopted in many domains, including CSI feedback.
The FLOP numbers of the encoder at the UE and the decoder at the BS for the two-sided NN model must be evaluated.
The BS has better computational power than the UE.
Accordingly, the NN model at the UE must be lightweight.
Although the number of FLOPs is the main KPI of the computational complexity of NNs, it does not fully reflect the inference speed of NNs in computing units.
The memory access cost and the degree of parallelism are not considered in the FLOP number.
In terms of the NNs with the same FLOP number, the one with less memory access and a higher degree of parallelism can be much faster than the one with more memory cost and a lower degree of parallelism.
Consequently, the FLOP number is the indirect KPI of NN inference speed, and the discrepancy between the FLOP number and the NN inference speed is unavoidable.
The memory access cost and the degree of parallelism must also be considered in the NN speed evaluation. 

The memory storage of the NN model must also be evaluated.
Two parts must be considered when assessing the memory storage occupy, namely, the NN model size and the NN weights.
The quantization bit of NN weights also affects the storage requirement.
For example, the binary weights need less storage than the 32-bit floating point ones.

\subsubsection{Generalization of the AI-enabled CSI Feedback}
\label{genera}
{{
As mentioned in Section \ref{reasonWhy}, the excellent performance of AI-enabled CSI feedback can be attributed to the ability of learning knowledge from the data through an end-to-end approach.
The NN model can be regarded as overfitting in a certain distribution of the training dataset.}}
If the distribution of the training dataset is the same as (or near) that of the deployment scenario, then the trained NN model efficiently works.
However, the performance unavoidably drops if the model is deployed to the systems with a different distribution, resulting in an NN generalization problem.
The key problem that must be clarified and solved in NN generalization is assessing and improving the generalization ability of the feedback NN model.

The common evaluation method is to assess the NN model with the test dataset, where the channel model (such as CDL-A) is different from that (such as CDL-C) adopted in the training dataset.
The generalization performance heavily depends on the generated test dataset.
An NN model may perform rather differently in a different test dataset.
For example, if the distributions of two different channel models are similar, then the performance drop may be very small.
Therefore, a public and general test dataset is essential.

Two strategies are expected to improve the generalization of the CSI feedback NN model, namely, training dataset mix and online learning.
The generalization errors in the feedback NN model result from the mismatch between the distributions of the training and test datasets.
If the training dataset covers the distribution of the test dataset, then the generalization errors are small.
Therefore, generating a training dataset that covers the entire distribution can considerably improve the generalization ability of the feedback NN model.
The mixed dataset-based strategy is developed based on this observation.
The CSI samples generated with different channel models are mixed to form a large training dataset.
However, the mixed training dataset cannot cover all distributions.
If a new distribution appears, then online learning is required.

Online learning involves not only the air interface but also the upper layer.
The protocol of online learning for the one-sided or two-sided feedback NN models should be carefully designed.
The protocol needs to define when to conduct online learning, how to collect CSI samples, which UE must participate in the online learning, how to allocate the computational power, and so on.
The protocol for online learning is a general challenge in an AI-native air interface.
{
In addition, more advanced learning tools, such as transfer learning and meta-learning, should be adopted to accelerate the online learning and reduce the training overhead.
In transfer learning, the NNs are finetuned with new CSI samples rather than trained from scratch.
Meta-learning provides a better initialization of the NN weights and helps the NNs quickly adapt to a new task (distribution), thereby considerably reducing the training epochs and the requirement for online training samples \cite{9413598}.
}

%In addition, the configurations of the systems, including the antenna number and arrangement and the subcarrier number, are not fixed.
%Therefore, the trained NN model of CSI feedback also needs to be generalizable (flexible) to different configurations. 

\subsubsection{Collaboration between the UE and the BS of the AI-enabled CSI Feedback}
\label{SCollaboration}

In 5G NR, all UE and BSs share the same codebook, and the extra collaboration between the BS and the UE is not needed.
However, collaboration is indispensable in AI-enabled CSI feedback.
We divide the collaboration into three types, as shown in Fig. \ref{Collaboration}.
\begin{itemize}
    \item {\emph{No collaboration:}}
    In this case, the AI model is deployed at one side (only UE or BS), and few standardization impacts are observed.
    The AI-enabled one-sided refinement-based framework in Section \ref{f1} is a representative use case of this type.
    However, in all AI-enabled feedback algorithms, the NN model is trained with the collected CSI samples.
    Some studies proposed to train the feedback NNs with the uplink CSI samples on the basis of the assumption that the bidirectional channels share the same distribution.
    However, this training strategy has not been verified in practical systems.
    If the distribution reciprocity does not hold, then the CSI dataset should be transmitted from the UE to the BS, where collaboration is necessary.

    \item {\emph{Signaling-based collaboration without NN model transfer:}} This type of collaboration mainly includes training dataset collection and signaling-based NN model management.
    As mentioned in the first type, signaling-based collaboration is required if the UE needs to transmit the stored CSI samples to the BS.
    In addition, the UE may store several NN models for different scenarios, and the BS needs to send some signals to the UE for NN model switching or activation.
    
    \item {\emph{Signaling-based collaboration with NN model transfer:}}
    This collaboration includes all collaboration operations in the second mechanism.
    The difference between them is that the NN model in the last type must be transmitted among the BS and the UE.
    Given the dataset privacy, the NN model for the CSI feedback may need to be trained with federated learning or model splitting.
    Federated learning is a decentralized learning framework, where the UE upload the local models to the server at the BS, and the BS aggregates the received NN model and then transmits the aggregated NN model to the UE.
    In model splitting, the encoder and the decoder are trained at the UE and the BS, respectively, and the gradients are transmitted between the BS and the UE.
    In addition, the BS may train a better NN model because more CSI samples are collected after deployment.
    The new NN model must be transmitted to all the UE in the cell.
\end{itemize}

{
The above-mentioned three collaboration mechanisms can be mixed and used according to actual communication scenarios.}

\subsubsection{Shared Information of the AI-enabled CSI Feedback}
Intellectual property is essential in standardization.
The company, whose technology is agreed in the 3GPP and deployed to practical systems, can earn huge economic profits from the technology.
Therefore, the 3GPP must clarify the content that should be standardized in AI-enabled CSI feedback.
As agreed by most companies, the NN architecture does not need to be standardized, and each company can design and train a feedback NN model with their own dataset.

The one-sided NN-based CSI feedback at the BS (UE) can work well even if the other side, that is, the UE (BS), has no knowledge of the NN model. 
However, the encoder at the UE and the decoder at the BS should be trained together for the two-sided autoencoder-based CSI feedback.
If the encoder and the decoder do not match, then the feedback accuracy will considerably drop, resulting in a great decrease in system throughput.
In image compression, some recent studies have attempted to improve the compression performance by designing the encoder (compression module) without changing the original JPEG-2000 decoder.
Their experimental results show that the performance can be considerably improved only when the encoder can be edited.
On this basis, the manufacturers of the BSs can release their decoder, and the manufacturers of the UE can design the encoder on the basis of the released decoder.
The UE requires storing several encoders and switching the encoder according to the BS types.
However, the model download is needed if the decoder at the BS changes, as mentioned in Section \ref{SCollaboration}.

\subsubsection{Joint Design with Channel Prediction}

  \begin{figure}[t]
	\centering 
	\includegraphics[width=0.48\textwidth]{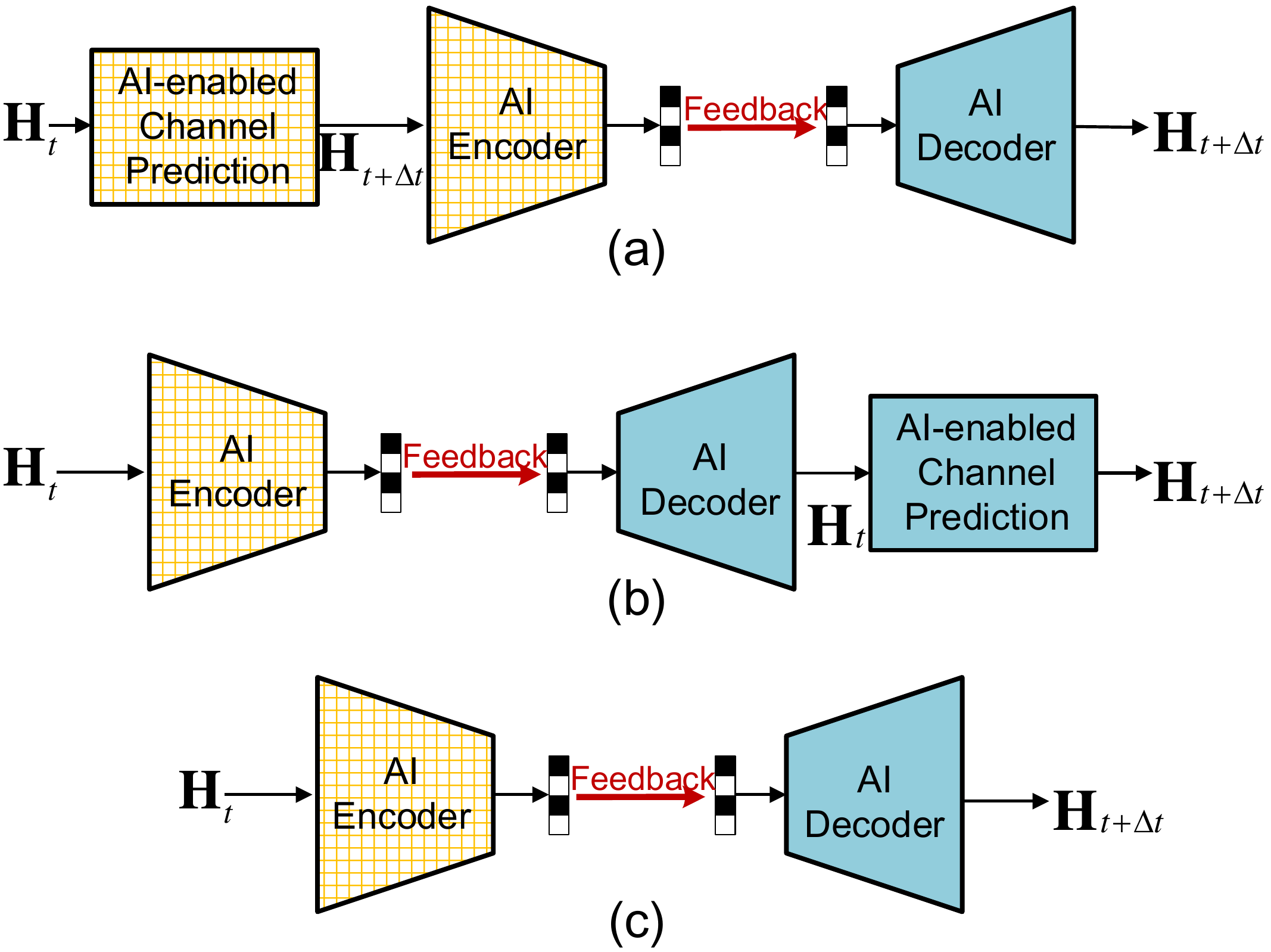}
	\caption{\label{FP}{Three frameworks for joint AI-enabled CSI feedback and prediction}. The first two NN-based prediction modules work at the UE and the BS, and the last one implicitly realizes channel prediction.}  
 \vspace{-0.75cm}
\end{figure}

As mentioned in Section \ref{scope3}, channel prediction is being studied in Release 18 because of the channel aging in the high-speed scenario.
Although the channel rapidly changes, the correlation among the adjacent CSI still holds.
The change within a short time can be predicted.
Recently, AI-based channel prediction has been explored in some studies.
{If the difference between the predicted and estimated CSI at the UE is tolerable, then feedback is no longer needed. 
The BS uses the predicted CSI to design precoding vectors.
In this case, the BS and the UE share the same CSI predictor.}
If the difference is large, three types of design will emerge when the feedback and prediction are jointly built as shown in Fig. \ref{FP}.
We use the AI-enabled two-sided explicit CSI feedback as an example.
In the first joint design frameworks, the UE first predicts the future channel, ${\bf H}_{t+\Delta t}$, based on the current channel, ${\bf H}_{t}$.
Then, the predicted channel is fed back to the BS by an autoencoder.
In the second one, the current channel, ${\bf H}_{t}$, is directly fed back to the BS by an autoencoder.
The reconstructed channel, ${\hat {\bf H}}_t$, is then used as input to a channel prediction module to reduce the effects of channel aging.
The above-mentioned frameworks jointly design the feedback and prediction explicitly, where two modules, namely, channel feedback and prediction modules, are adopted.
The last framework jointly designs the channel feedback and the prediction implicitly.
This framework seems to be the same as that in Fig. \ref{ThreeFramework}(c).
The main difference is the training phase, where the input is the current channel, and the output of the autoencoder (decoder) is the future channel.
The framework embeds channel prediction into the feedback process.

The main advantage of the first framework is that the UE also obtains the future CSI, which can be used in the design of the receiver.
However, the extra module increases the computational complexity at the UE.
In the second framework, the increase in complexity can be ignored because of the large computational power at the BS.
No extra modules are added in the third framework, and the main concern is the lack of interpretability.
The practical system must select the most suitable framework according to the application scenario.

\subsubsection{Combination with the Enhanced Reciprocity-based Feedback }
%In comparison with the Type II codebook in 3GPP Release 15, the codebook in 3GPP Release 16 has been enhanced by allowing an improved performance-overhead tradeoff. 
In 3GPP Release 17, the exploitation of the angle-delay reciprocity between the uplink and downlink channels is considered and discussed to improve the performance (including the accuracy and overhead) of the enhanced Type II codebook in 3GPP Release 16 \cite{3gppr17}.
Given that the angle-delay information of the bidirectional channels is reciprocal, the UE need not feedback this information to the BS, thereby reducing feedback overhead and computational complexity.

Accordingly, AI-enabled CSI feedback enhancement also requires studying the angular-delay reciprocity.
The key problem is introducing the uplink channel information to the downlink CSI reconstruction at the BS, thereby reducing the feedback of the information shared by the bidirectional channels.
Meanwhile, channel estimation is the basis of the CSI feedback.
If less information of the downlink channel must be fed back, then the channel estimation does not estimate the information that is not fed back.
Therefore, the pilot overhead can be considerably reduced if the channel estimation and feedback are jointly designed on the observation of the angular-delay reciprocity.

 {
\subsubsection{Others }
During deployment, more challenges and problems must to be addresses.
First, some new use cases, such as reconfigurable intelligent surfaces, are expected to play an important role in the future.
Some new characteristics may appear in these cases.
The combination of AI-enabled CSI feedback with these new cases is worth exploring.
Second, the computational powers of different UE or a specific UE at different times vary greatly.
Therefore, feedback NNs should be scalable with different complexities instead of training and storing many NNs with various complexities.
Third, the performance of AI-enabled CSI feedback should be evaluated with the CSI samples collected from the practical systems before deployment.

}

\section{Conclusions}
\label{future}
In this article, an overview of the AI for CSI feedback enhanced in 5G-Advanced is provided.
{
First, non-AI CSI feedback frameworks
%, including the codebook- and CS-based frameworks, 
and three representative frameworks of AI-enabled CSI feedback enhancement
%, including one-sided implicit feedback refinement, two-sided autoencoder-based implicit feedback, and two-sided autoencoder-based explicit feedback, 
are introduced and compared.
Then, the scope of AI for CSI feedback enhancement
%, including overhead reduction, accuracy improvement, and channel prediction, 
is introduced and discussed.
Finally, the considerations in the standardization of AI for CSI feedback enhancement, especially
focusing on accuracy, complexity, collaboration, generalization, information sharing, joint design with channel prediction, and reciprocity, are elaborated.}

As pointed out in \cite{3gpp599}, the goal of studying three representative use cases is to lay the foundation for future air interface use cases leveraging AI techniques.
{
The above-mentioned challenges will be addressed by introducing advanced AI techniques and communication expert knowledge.}
The thorough study of AI for CSI feedback enhancement will guide the development of AI-native air interface in 5G-Advanced and 6G.

\bibliographystyle{IEEEtran}
\bibliography{magazine}

\end{document}